\documentclass[%
reprint,
superscriptaddress,
altaffilletter,
showpacs,
showkeys,
amsmath, amssymb, aps, prl, floatfix,
]{revtex4-1}

\usepackage{graphicx}
\usepackage{dcolumn}
\usepackage{bm}
\usepackage{comment}
\usepackage[usenames,dvipsnames]{color}
\usepackage{hyperref}
\hypersetup{colorlinks = true, allcolors = blue}
\usepackage[mathlines]{lineno}
\usepackage{multirow}
\DeclareGraphicsExtensions{.pdf,.png,.jpg}

\newcommand{\onu}{$0 \mathrm{\nu\beta \beta}$ }
\newcommand{\tnu}{$2 \mathrm{\nu\beta \beta}$ }
\newcommand{\TL}{$^{208}\mathrm{Tl}$}

\newcommand{\THO}{$^{232}\mathrm{Th}$}

\newcommand{\al}{$\alpha$}
\newcommand{\be}{$\beta$}
\newcommand{\ga}{$\gamma$}
\newcommand{\QBB}{$\mathrm{Q}_{\beta\beta}$}

\newcommand{\ckky}{counts/(keV kg yr)}

\begin{document}

\title{First Result on the Neutrinoless Double Beta Decay of $^{82}$Se with  CUPID-0}

\newcommand{\sapienza}{\affiliation{Dipartimento di Fisica, Sapienza Universit\`a di Roma, P.le Aldo Moro 2, 00185, Roma, Italy}}
\newcommand{\infnroma}{\affiliation{INFN, Sezione di Roma, P.le Aldo Moro 2, 00185, Roma, Italy}}
\newcommand{\lnl}{\affiliation{INFN  Laboratori Nazionali di Legnaro, I-35020 Legnaro (Pd) - Italy}}
\newcommand{\lngs}{\affiliation{INFN  Laboratori Nazionali del Gran Sasso, I-67100 Assergi (AQ) - Italy}}
\newcommand{\lbl}{\affiliation{Lawrence Berkeley National Laboratory , Berkeley, California 94720, USA}}
\newcommand{\infnge}{\affiliation{INFN  Sezione di Genova, I-16146 Genova - Italy}}
\newcommand{\unige}{\affiliation{Dipartimento di Fisica, Universit\`{a} di Genova, I-16146 Genova - Italy}}
\newcommand{\infnmib}{\affiliation{INFN  Sezione di Milano Bicocca, I-20126 Milano - Italy}}
\newcommand{\unimib}{\affiliation{Dipartimento di Fisica, Universit\`{a} di Milano Bicocca, I-20126 Milano - Italy}}
\newcommand{\csnsm}{\affiliation{CNRS/CSNSM, Centre de Sciences Nucl$\acute{e}$aires et de Sciences de la Mati$\grave{e}$re, 91405 Orsay, France}}
\newcommand{\cea}{\affiliation{IRFU, CEA, Universit$\acute{e}$ Paris-Saclay, F-91191 Gif-sur-Yvette, France}}
\newcommand{\gssi}{\affiliation{Gran Sasso Science Institute, 67100, L'Aquila - Italy}}
\newcommand{\usc}{\affiliation{Department of Physics  and Astronomy, University of South Carolina, Columbia, SC 29208 - USA}}
\newcommand{\mpi}{\affiliation{Max-Planck-Institut für Physik, D-80805 München, Germany}}
\newcommand{\dis}{\affiliation{DISAT, Universit\`a dell'Insubria, 22100 Como, Italy}}
\newcommand{\MIT}{\affiliation{Massachusetts Institute of Technology, Cambridge, MA 02139, USA}}

\author{O.~Azzolini}\lnl
\author{M.T.~Barrera}\lnl
\author{J.W.~Beeman}\lbl
\author{F.~Bellini}\email[Corresponding author: ]{fabio.bellini@roma1.infn.it}\sapienza\infnroma
\author{M.~Beretta}\unimib\infnmib
\author{M.~Biassoni}\infnmib
\author{C.~Brofferio}\unimib\infnmib
\author{C.~Bucci} \lngs
\author{L.~Canonica}\altaffiliation{Present address: Max-Planck-Institut für Physik, D-80805 München, Germany}\lngs\MIT
\author{S.~Capelli}\unimib\infnmib
\author{L.~Cardani}\infnroma
\author{P.~Carniti}\unimib\infnmib
\author{N.~Casali}\infnroma
\author{L.~Cassina}\unimib\infnmib
\author{M.~Clemenza}\unimib\infnmib
\author{O.~Cremonesi}\infnmib
\author{A.~Cruciani}\infnroma
\author{A.~D'Addabbo} \lngs
\author{I.~Dafinei}\infnroma
\author{S.~Di~Domizio}\unige\infnge
\author{F.~Ferroni}\sapienza\infnroma
\author{L.~Gironi}\unimib\infnmib
\author{A.~Giuliani}\csnsm\dis
\author{P.~Gorla} \lngs
\author{C.~Gotti}\unimib\infnmib
\author{G.~Keppel}\lnl
\author{L.~Marini}\unige\infnge
\author{M.~Martinez}\sapienza\infnroma
\author{S.~Morganti}\infnroma
\author{S.~Nagorny}\lngs\gssi
\author{M.~Nastasi}\unimib\infnmib
\author{S.~Nisi}\lngs
\author{C.~Nones}\cea
\author{D.~Orlandi}\lngs
\author{L.~Pagnanini}\lngs\gssi
\author{M.~Pallavicini}\unige\infnge
\author{V.~Palmieri}\lnl
\author{L.~Pattavina}\lngs\gssi
\author{M.~Pavan}\unimib\infnmib
\author{G.~Pessina}\infnmib
\author{V.~Pettinacci}\infnroma
\author{S.~Pirro}\lngs
\author{S.~Pozzi}\unimib\infnmib
\author{E.~Previtali}\infnmib
\author{A.~Puiu}\unimib\infnmib
\author{F.~Reindl}\altaffiliation{Present address: Institut für Hochenergiephysik der ÖAW, A-1050 Wien, Austria. Atominstitut, Technical University Vienna, A-1020 Wien, Austria} \infnroma
\author{C.~Rusconi}\lngs\usc
\author{K.~Sch\"affner}\lngs\gssi
\author{C.~Tomei}\infnroma
\author{M.~Vignati}\infnroma
\author{A.~S.~Zolotarova}\cea 

\date{\today}

\begin{abstract}
We report the result of the  search for neutrinoless double beta decay of $^{82}$Se obtained with CUPID-0,  the first large array of scintillating Zn$^{82}$Se cryogenic calorimeters implementing  particle identification. We observe  no signal in a 1.83 kg yr $^{82}$Se exposure and we set the most stringent lower limit on  the \onu $^{82}$Se half-life T$^{0\nu}_{1/2}>$ 2.4$\times \mathrm{10}^{24}$ yr (90\% credible interval), which corresponds  to an effective Majorana neutrino mass  m$_{\beta\beta} <$ (376-770) meV depending on the nuclear matrix element calculations. The heat-light readout provides a powerful tool for the rejection of \al\ particles and allows us to suppress the background in the region of interest down to  (3.6$^{+1.9}_{-1.4}$)$\times$10$^{-3}$\ckky, an  unprecedented level for this technique. 
\end{abstract}

\maketitle
The simultaneous occurrence of two beta decays  ($2 \mathrm{\nu\beta \beta}$) \cite{PhysRev.48.512} is a nuclear transition observable in a total of 35 even-even nuclei for which the sequence of two single beta decays is energetically forbidden or highly spin suppressed. This transition has so far been measured  for 11 nuclei \cite{Barabash:2015eza}.
The $\beta\beta$ process without neutrino emission ($0 \mathrm{\nu\beta \beta}$) \cite{Furry} is predicted in several extensions of the standard model  of particle physics in which neutrinos are their own antiparticles \cite{giuntibook}. Its discovery would establish the total lepton number nonconservation and the nature of neutrinos as Majorana fermions \cite{PhysRevD.25.2951}, providing support to leptogenesis theories \cite{lepto}. 
In the case of the light Majorana neutrino exchange model,  the \onu decay rate is proportional to  the square of the effective Majorana neutrino mass m$_{\beta\beta}$=$|\sum_{i}$U$^2_{ei}$m$_i|$, where U$_{ei}$ are elements of the neutrino mixing matrix, and m$_i$ are the neutrino mass eigenvalues;  hence its measurement would also constrain the  neutrino mass scale. \\
The  signature of the \onu decay is a peak in the summed energy spectrum of the electrons at the transition energy \QBB, that must be identified  in presence of environmental background \cite{Cremonesi:2013vla}. The sensitivity of an experiment is, therefore, determined by the number of  $\beta\beta$ emitting isotopes, the energy resolution and the background level at \QBB.
At present, no \onu evidence has been found and actual limits on the half-life lie in the range of (10$^{24}$-10$^{26}$) yr \cite{Cremonesi:2013vla,DellOro:2016tmg, Albert:2014awa, Albert:2017owj,KamLAND-Zen:2016pfg,Aalseth:2017btx, Alduino:2017ehq,Agostini:2017iyd,Agostini:2018tnm}.\\
 Among the experimental techniques employed in this research field, cryogenic calorimeters (usually called bolometers) play a leading role \cite{pirroreview}. In such devices, a highly sensitive thermometer measures the temperature rise induced in a crystal by a particle interaction \cite{bolo-review1,bolo-review2}.
 This technology, thanks to the wide choice of crystal compounds, allows us to embed the \onu source in the detector itself. Moreover, it features  excellent energy resolution and  very high  detection efficiency.
 The CUORE experiment \cite{Artusa:2014lgv}  recently demonstrated \cite{Alduino:2017ehq} that a detector composed of a 1000 individual bolometers can be successfully used for the study of the \onu decay of $^{130}$Te (\QBB$\sim$2527 keV \cite{Rahaman:2011zz}). The sensitivity of CUORE is mainly limited by energy-degraded $\alpha$ particles, emitted by surface contamination on the crystals and on the copper supporting structure \cite{Alduino:2017qet}.
The CUPID project (CUORE Upgrade  with Particle IDentification) \cite{Wang:2015raa,Wang:2015taa} aims to enhance the  sensitivity by 2 orders of magnitude and, thus, test the \onu  decay in the inverted hierarchy scenario of neutrino masses \cite{Artusa:2014wnl}. 
To reach this goal CUPID will increase the source mass and reduce the background by using isotopically enriched bolometers with active particle identification. This can be achieved by  a scintillating bolometer \cite{nsvecchioarticoloCaF2,Pirro:2005ar,Poda:2017jnl} in which a  small fraction of the released energy is converted into scintillation light that is absorbed by a disklike bolometer acting as light detector. The dual readout provides the ultimate tool for particle identification and background rejection \cite{Gironi:2009ay,Arnaboldi:2010jx,Beeman:2013vda,Beeman:2013sba,Artusa:2016maw}; the scintillation induced by  \al\ particles, in fact, is characterized by a different amplitude and time-development compared to isoenergetic electrons.
Moreover,  the flexibility in the choice of the detector material allows us to select  \onu decaying isotopes with a \QBB\ greater than 2615 keV,  the energy of the most intense natural high-energy   \TL\ \ga\  line, thus reducing the \ga\ background by about 1 order of magnitude \cite{Bucci:2009fk}.\\
In this Letter we report the results of CUPID-0, the first kg-scale CUPID demonstrator employing enriched scintillating bolometers for the study of the \onu decay of $^{82}$Se [\QBB=(2997.9$\pm$0.3) keV \cite{Lincoln:2012fq}]. 
The CUPID-0 detector is described in Ref.~\cite{Azzolini:2018tum}. 
The array consists of 24 Zn$^{82}$Se crystals 95$\%$ enriched in $^{82}$Se (total mass of 9.65 kg, corresponding to 5.13 kg of $^{82}$Se) and two natural ZnSe crystals (total mass of 0.85 kg, corresponding to 40\,g of $^{82}$Se). Details about the production of enriched Zn$^{82}$Se crystals can be found in Ref.~\cite{Dafinei:2017xpc}.
The data from two enriched crystals, not properly functioning, and from the two natural crystals are not considered in the current analysis. The  number of $^{82}$Se nuclei under investigation is, therefore, (3.41$\pm$0.03)$\times$10$^{25}$. 
Each Zn$^{82}$Se is held in a  copper frame by means of small polytetrafluoroethylene supports, side surrounded by a 3M Vikuiti plastic reflective foil to increase the light collection efficiency and monitored by a light detector (LD).  
The LD is a 170-$\mu$m-thick Ge disk \cite{Beeman:2013zva} coated on one side with a SiO 60-nm-thick layer  to enhance  light absorption \cite{Mancuso:2014paa}. 
Each device is equipped with a neutron trasmutation doped  Ge thermistor \cite{Haller}, biased with a constant current, and acting as temperature-voltage transducer. A P-doped Si Joule heater \cite{Andreotti:2012zz}, glued to each crystal, periodically injects a fixed amount of energy to equalize the bolometer response \cite{stabilization,Carniti:2017zkr}.
The front end electronics comprises an amplification stage,  a six-pole anti-aliasing active Bessel filter (120 dB/decade) and an 18 bit analog-to-digital converter board  operating at 1(2) kSPS for the Zn$^{82}$Se (LD).  
The detector is anchored to the mixing chamber of an Oxford 1000 $^3$He/$^4$He dilution refrigerator operating at a base temperature of about 10 mK, located  in Hall A of the Laboratori Nazionali del  Gran Sasso (average depth $\sim$3650 m water equivalent \cite{lngsdepth}).
The cryogenic system and electronics are detailed in Refs.~\cite{Pirro:2006mu,designcuore,carniti2016, Arnaboldi:2017aek, Arnaboldi:2006mx, Arnaboldi:2010zz, Arnaboldi:2015wvc}.

\begin{figure}[tb]
\centerline{\includegraphics[width=8.5cm]{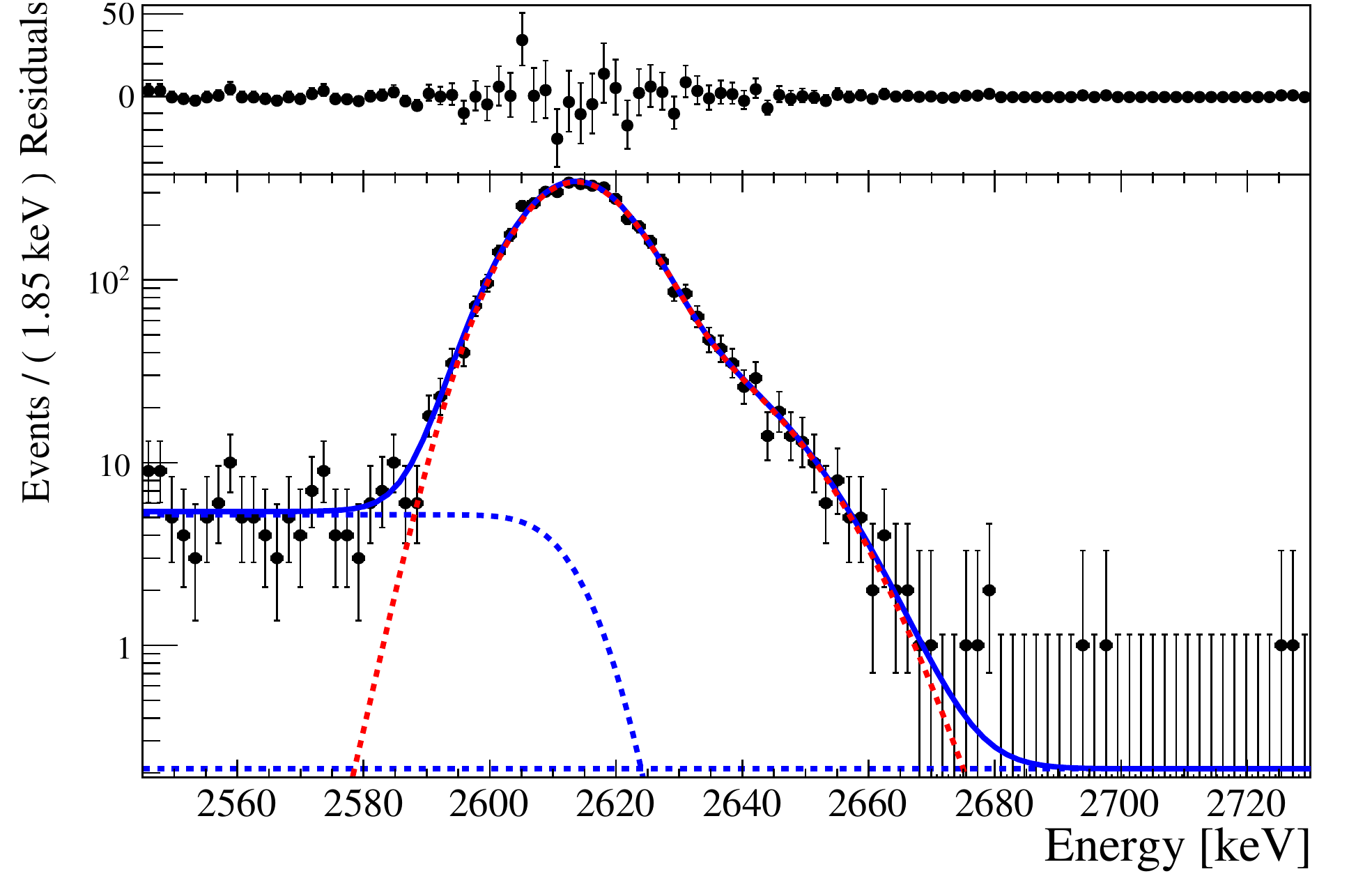}}
\caption{\TL\ \ga\ line in calibration data summed over all channels. The solid blue line is the result of the fit as described in the main text. The dashed red line represents the bi-Gaussian model for the \ga\ peak while the dashed blue lines are the flat and multi-Compton terms. Residuals between the data and best-fit model are shown in the upper panel.}
\label{model}
\end{figure}

\begin{figure}[htb]
\centerline{\includegraphics[width=8.5cm]{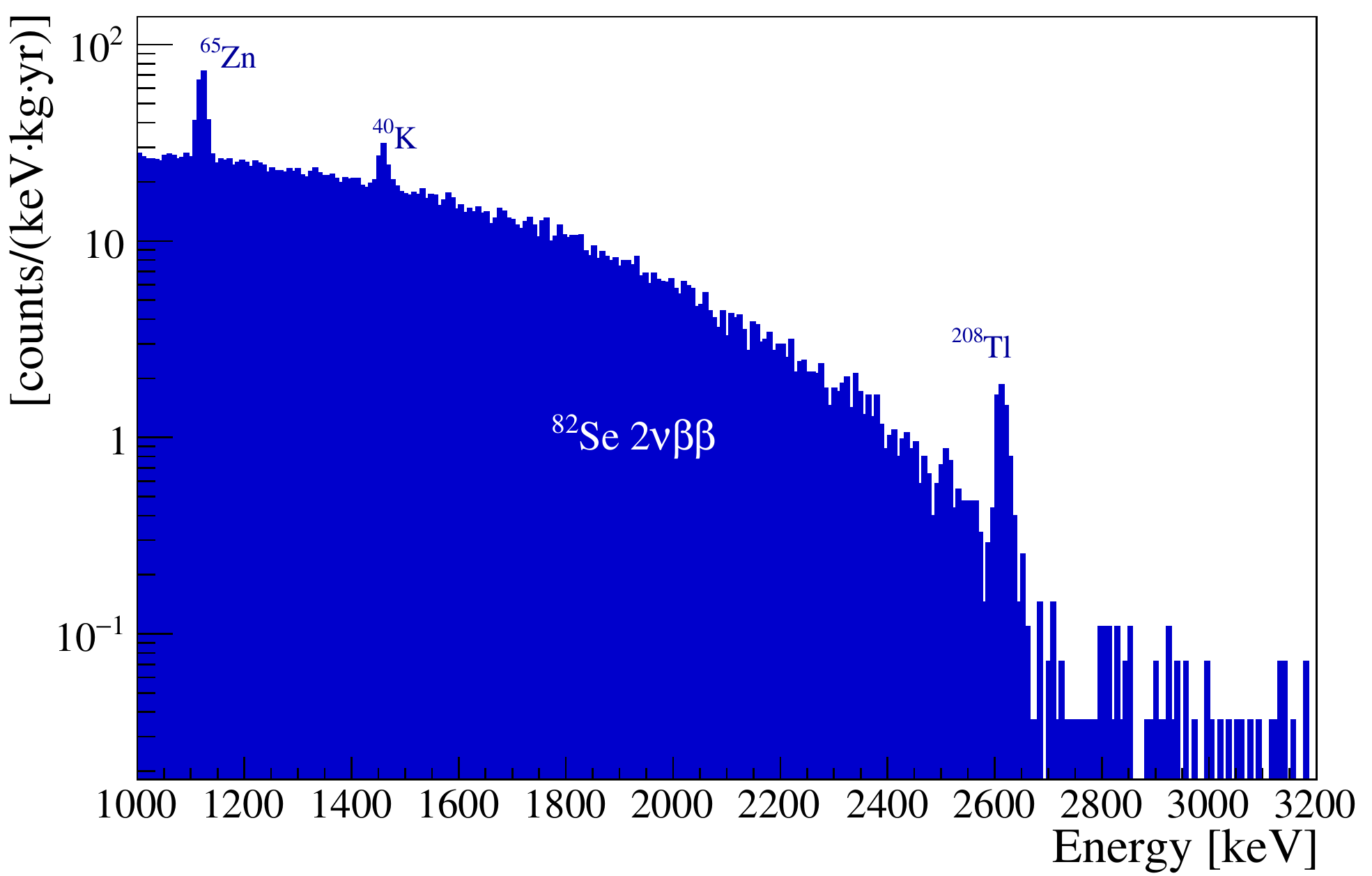}}
\caption{Single-hit reconstructed energy spectrum after the selection on the Zn$^{82}$Se thermal pulses.  Three prominent peaks are visible on  the top of the continuous spectrum generated by the \tnu decay of $^{82}$Se [$\tau_{1/2}$=(0.92 $\pm$ 0.07)$\times$10$^{20}$ yr \cite{Barabash:2015eza}]:  the cosmogenically activated $^{65}$Zn,  the  $^{40}$K and  \TL\  line due to natural radioactivity. The bin width  is 8 keV.} 
\label{spectrum}
\end{figure}

The data we present here were collected between June and December 2017 and are divided in four blocks called data sets. At the beginning and at the end of each data set we perform a calibration exposing the detector to thoriated wires which provide \ga\ lines up to 2615 keV. 
For each bolometer (i.e. Zn$^{82}$Se or LD) we acquire the complete data stream. We implement a software trigger on the Zn$^{82}$Se and search for a simultaneous signal on the LD. The trigger  threshold is channel dependent and ranges between 10 and 110 keV. Typical trigger rates are 2 mHz per bolometer (50 mHz in calibration). The heaters produce a pulse every 400 sec, that is automatically flagged by the data acquisition.
For each Zn$^{82}$Se waveform we analyze 4 sec after the trigger and 1 sec before (pretrigger). 
Average rise and decay times, defined as  (90 -10)\% time difference  of the leading edge and the  (30 -90)\% time difference  of the trailing edge, are 10 ms and 40 ms respectively.
We use the optimum filter technique \cite{Gatti:1986cw,Radeka:1966} to estimate  pulse height and pulse shape parameters. We build the signal waveform template averaging the physical events with energy between 1800 and 2700 keV and compute the noise power spectrum on  randomly chosen waveforms that do not contain triggered pulses.
We  correct the filtered  amplitude  for the shift in thermal gain due to temperature variations by using the constant heater pulse amplitude and the heater pretrigger baseline level as a proxy for the Zn$^{82}$Se temperature \cite{stabilization}. 
The intrinsic FWHM energy resolution, determined by noise fluctuations at the filter output, is on average $\sim$ 5 keV.\\
We use the \THO\ calibration data to derive the Zn$^{82}$Se amplitude-to-energy conversion and the detector response function. 
We fit the positions of the  most intense  $\gamma$ lines in the range (511-2615) keV with a zero intercept parabolic function. The extrapolation of the calibration function at the \QBB\ energy gives an uncertainty  $\sigma_{Q_{\beta\beta}}$= 3 keV. The distribution of the residuals as a function of the energy is flat with a weighted average of (0.42$\pm$0.05) keV.   We do not correct for this small offset at \QBB\ and we treat $\sigma_{Q_{\beta\beta}}$ as a  systematic uncertainty. 
We use the 2615 keV line from the  \TL\ line as a proxy for the detector response function for the \onu event. 
We parametrize its line shape with a double Gaussian $\mathcal{G}$($\mu_{p}$, $\sigma_{p}$, $\rho$, $\eta$, $\epsilon$) where $\mu_p$ and $\sigma_p$ are the mean and width of the primary peak, and $\rho$, $\eta$, $\epsilon$ are the ratio of the mean, width, amplitude of the secondary to the primary peak respectively.  This is the simplest model which well reproduces the detector response function of the observed peaks over the entire spectrum. Deviations from the single Gaussian model were already observed in other bolometric experiments \cite{Alfonso:2015wka,Artusa:2014lgv,Alduino:2016zrl}. 
We estimate the five parameters with an unbinned extended maximum likelihood (UEML) fit, including  a term to model the multi-Compton continuum and a flat background (Fig.~\ref{model}).
We fit the line shape to the other prominent calibration peaks  with $\rho$, $\eta$, $\epsilon$  constrained to the inferred values at the \TL\ line.
We linearly extrapolate $\sigma_p$ at \QBB; the exposure-weighted harmonic mean FWHM energy resolution results to be  (23.0$\pm$0.6) keV.\\
We  select \onu candidates applying  the following criteria on the Zn$^{82}$Se thermal pulses only.
We first reject periods of detector instability due to electronics or cryostat malfunctions, reducing the exposure by less than 1\%. 
We require only one pulse  in a triggered window and the waveform to be consistent with the signal template. To this end we use six different pulse shape parameters normalized over the energy spectrum, in order to have a constant efficiency.
The normalization is performed using \ga s produced in a dedicated run with an AmBe neutron source;  neutron reactions in the detector and in the surrounding structure generate a continuum of  \ga\ s  up to several MeVs.
We maximize a score function defined as the ratio of the signal efficiency to the square root of the off-peak background efficiency using  50\% of randomly selected 1115 keV $^{65}$Zn \ga\ peak events \cite{Alduino:2016zrl}. $^{65}$Zn is a short-lived isotope (T$_{1/2}$=244 d) produced  by cosmogenic activation of the Zn powder and represents the most intense line visible in the spectrum with a rate of a few counts/(kg d). 
To reduce the background  from events producing a signal in multiple crystals (as the ones induced by multiple Compton \ga's) we discard events on different bolometers if they occur within 20 ms (multiple-hit events). We gauge the time window on a selected sample of  \TL -induced double coincidence events whose summed energy is  2615 keV.  The energy spectrum of selected events  is shown in Fig \ref{spectrum}.

\begin{figure}[htb]
\centerline{\includegraphics[width=8.5cm]{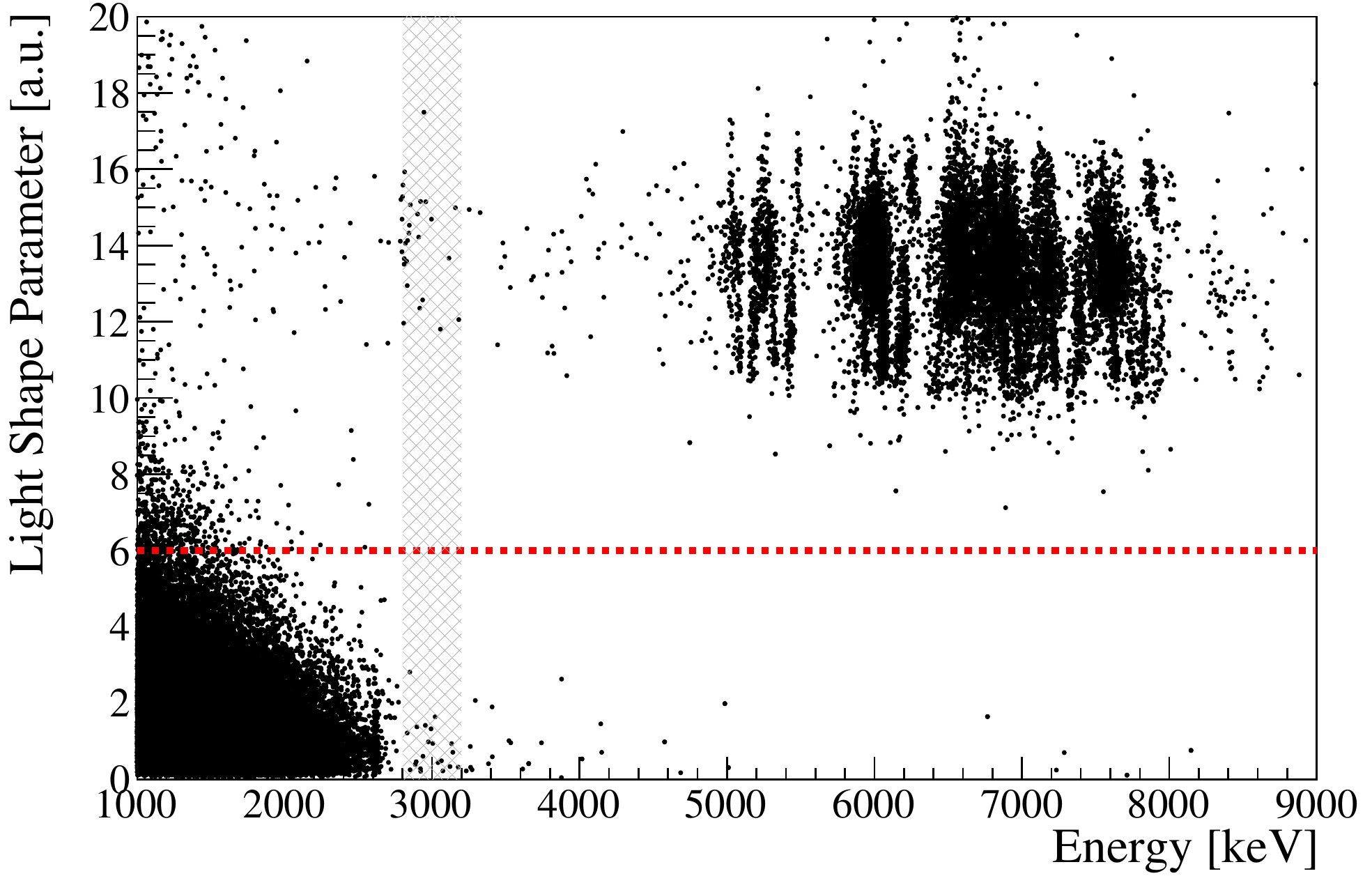}}
\caption{Single-hit events in the Light Shape Parameter-Energy plane. The \al\ events are concentrated in the right upper region; \be/\ga\ events populate the left lower corner. The red dotted  line indicates the acceptance threshold. The shaded vertical band represents the analysis window used for the background evaluation at \QBB. The \THO\ calibration  applied to energy deposits induced by \al\ decays results in a 25\% positive shift compared to the nominal  Q$_{\alpha}$ transition energies.}
\label{scatter}
\end{figure}

\begin{figure}[htb]
\centerline{\includegraphics[width=8.5cm]{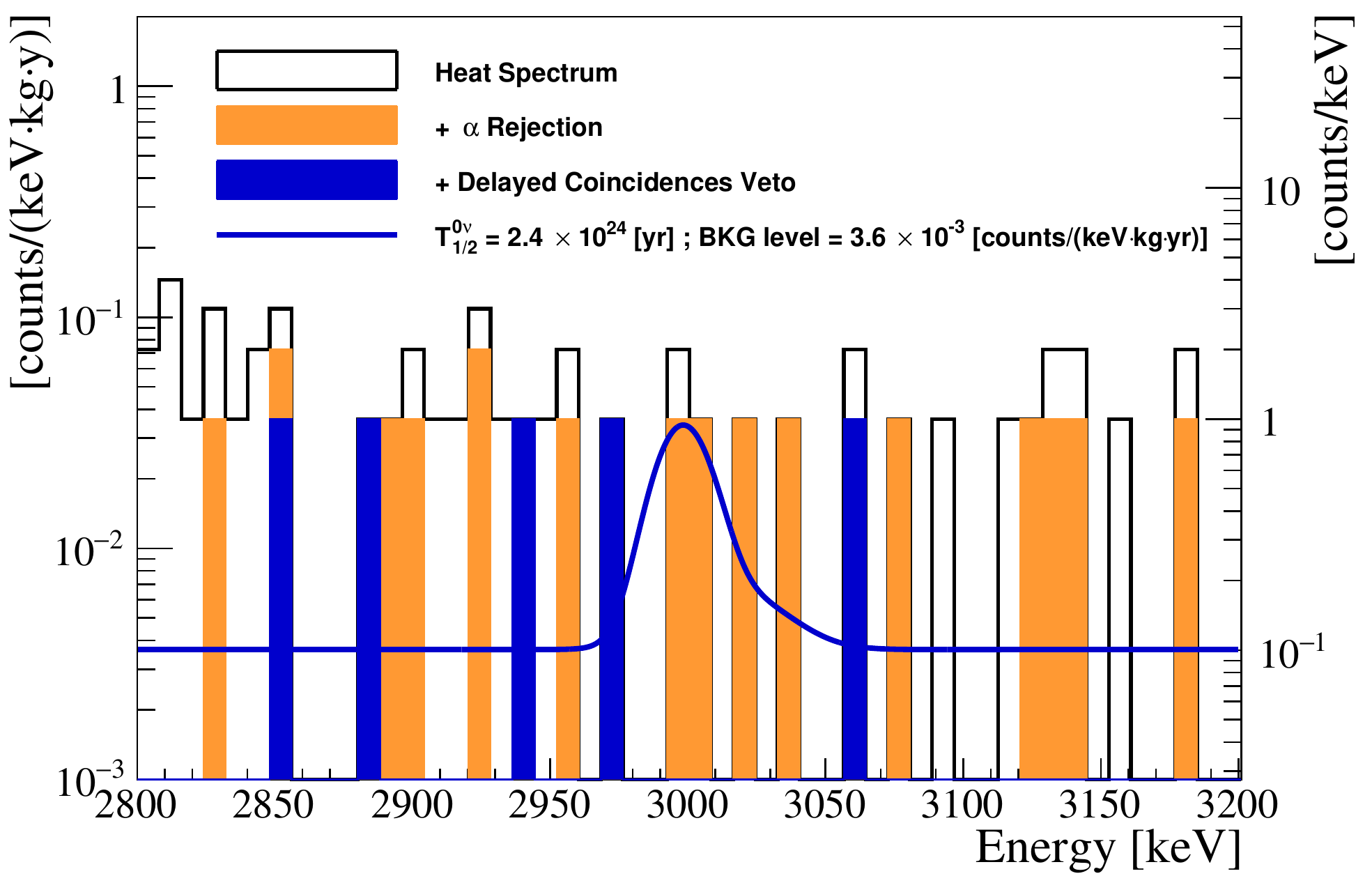}}
\caption{The energy spectrum in the analysis window. The open histogram is the spectrum obtained with the selection on Zn$^{82}$Se thermal pulses. The filled orange histogram includes the \al\ rejection through the LD pulse shape. The filled blue histogram is the final spectrum after the delayed-coincidences veto is  applied. The blue line is the fitted spectrum together with a hypothetical signal corresponding  to the 90\% credible interval (C.I.) limit of  T$^{0\nu}_{1/2}$=2.4$\times$10$^{24}$ yr. The detector response function is described in the main text. The bin width is 8 keV.}
\label{final}
\end{figure}

We then exploit the information of the LD. We estimate  the LD signal amplitude at a fixed time delay with respect to the Zn$^{82}$Se signal as described in  Ref.~\cite{Piperno:2011fp}.  
Typical rise and decay times are 4 and 8 ms respectively.  We use a shape parameter computed on the filtered light pulse as defined in Ref.~\cite{Artusa:2016maw} to reject \al\ particles. 
To build a control sample of high-energy \ga's we require multiple-hit events with multiplicity greater than 4 and a light signal amplitude incompatible with direct muon ionization. Such events are generated only by electromagnetic showers induced by the passage of a high-energy cosmic muon in the surrounding shields and their energy extends up to $\sim$ 5 MeV.  We select \be/\ga\ events with 100\% efficiency.  
The acceptance threshold and \al\ discrimination capability for single-hit events are shown in Fig.~\ref{scatter}.  From a Gaussian fit to the LD shape parameter distribution for events with energy greater than 4500 keV we obtain an \al\ misidentification probability lower than 10$^{-6}$.\\
We implement delayed coincidences to suppress the background induced by the internal \TL\ \be/\be +\ga\ decay from the  \THO\ chain.  
The $^{212}$Bi \al decays to \TL\ (Q$_{\alpha}$=6207 keV), which, in turn, \be decays to the stable isotope $^{208}$Pb with  Q$_{\beta}$=5001 keV and a half-life $\tau_{1/2}$=3.01 min. We veto any event succeeding a primary $^{212}$Bi \al\ event in a window corresponding to three times the half-life. If the contamination is close to the surface and the \al\ escapes the crystal, only part of the energy of the parent decay is collected. To identify such events we require the pulse shape of the primary event to be consistent with the reference  \al\ shape and the energy to be in the range (2.0-6.5) MeV. 
Figure.~\ref{final} shows the effect of the selection criteria on the energy spectrum in the analysis window selected for the background evaluation. This is approximately symmetric around \QBB\  and ranges between 2800 and 3200 keV, where we do not  expect peaking background contributions. 

The signal efficiency comprises: the probability that a \onu event is confined inside a single crystal, that is triggered and its energy properly reconstructed, and finally that it survives the selection criteria. We determine the probability of a \onu to be fully contained in a single crystal from GEANT4 simulation to be (81.0$\pm$0.2)\%.
We compute the other efficiencies for each data set and we quote the average weighted on the data set exposure.
We evaluate the trigger efficiency as the ratio of triggered to flagged heater pulses and the energy reconstruction efficiency as the probability of the monoenergetic heater pulse to be reconstructed within 3 Gaussian standard deviations \cite{Andreotti:2010vj}. The combined efficiency  is (99.44$\pm$0.01)\%.
Finally, we estimate the selection efficiency from a simultaneous fit on both the spectra of accepted and rejected events in the 1115 keV $^{65}$Zn peak in the sample not used for the optimization \cite{Andreotti:2010vj}. We sum all channels due to the limited statistics and derive a selection efficiency of (93$\pm$2)\%. 
We cross-check the selection efficiency  as a function of the energy selecting double-hit events; these are very likely a sample of true particle events since spurious coincidences are negligible. The ratio of events before and after applying the selection criteria is compatible with the efficiency computed on the peak of $^{65}$Zn in a range up to 2.6 MeV.
The total signal efficiency is, therefore, (75$\pm$2)\%.
We estimate the number of \onu candidates and the background index (BI) from a simultaneous UEML fit in the analysis window. For each data set the fit comprises the bi-Gaussian line shape for the \onu signal, with primary peak position fixed at Q$_{\beta\beta}$, and a flat  background component. The efficiency and the energy resolution are data set dependent. The decay rate $\Gamma^{0\nu}$  and the BI are treated as free parameters common to all the detectors and data sets. In a 1.83 kg yr $^{82}$Se (3.44 kg yr Zn$^{82}$Se) exposure we find no signal evidence and a BI=(3.6$^{+1.9}_{-1.4}$)$\times$10$^{-3}$\ckky,~consistent with the five observed events in the 400 keV analysis window.  

We estimate the systematics due to the uncertainty on the absolute energy scale,  the detector response function,  the efficiency and the exposure.
For each influence  parameter  we weight the likelihood with a Gaussian probability density function with the mean and width fixed to the best estimated  values and uncertainties respectively. We then integrate the likelihood  via numerical integration.\\
We set a  90\% C.I. Bayesian upper limit on $\Gamma^{0\nu}$ using a uniform prior in the physical region of $\Gamma^{0\nu}$ and marginalizing over the BI nuisance parameter: $\Gamma^{0\nu}<0.285\times$10$^{-24}$ yr$^{-1}$. This corresponds to a lower limit on the  half-life  of 
\begin{equation}
\nonumber
\mathrm{T^{0\nu}_{1/2}>2.4\times10^{24} \ yr\ (90\%~C.I.)}. 
\end{equation}
We evaluate the median 90\% C.I. lower limit sensitivity  from toy MC experiments to be:  T$^{0\nu}_{1/2}>$ 2.3$\times$10$^{24}$ yr. With the accumulate exposure, the probability to obtain a limit greater than the one we report in this paper is 44\%.
The CUPID-0 result surpasses by almost 1 order of magnitude the previous limit  of T$^{0\nu}_{1/2}>$ 3.6$\times$10$^{23}$ yr  \cite{Barabash:2010bd} obtained by NEMO  with a larger $^{82}$Se exposure ($\sim$3.5 kg yr).\\
In the light Majorana neutrino exchange model for the \onu decay, the effective neutrino mass m$_{\beta\beta}$ is related to T$^{0\nu}_{1/2}$ by
\begin{equation}
   \mathrm{(T^{0\nu}_{1/2})^{-1}  = G_{0\nu} \, |\mathcal{M}_{0\nu}|^2 \, m_{\beta\beta}^2}
   \label{eq:t12tombb}
\end{equation}

where G$_{0\nu}$ and M$_{0\nu}$ are the phase space factor of the decay and the dimensionless nuclear matrix element (NME). Using  G$_{0\nu}$  from Refs.~\cite{Kotila:2012zza,Stoica:2013lka}, the NME from Refs.~\cite{Engel:2016xgb,Yao:2014uta,Menendez:2008jp,Simkovic:2013qiy,Rodriguez:2010mn,Meroni:2012qf} and an axial coupling constant g$_a$=1.269 we find an upper limit on m$_{\beta\beta}<~$(376-770) meV. Despite the small CUPID-0 scale, the achieved m$_{\beta\beta}$  approaches the range of the most sensitive experiments in the field \cite{Albert:2014awa, Albert:2017owj,KamLAND-Zen:2016pfg,Aalseth:2017btx, Alduino:2017ehq,Agostini:2017iyd,Agostini:2018tnm}. 

In summary we find no evidence of $^{82}$Se \onu decay in 1.83 kg yr $^{82}$Se (3.44 kg yr Zn$^{82}$Se) exposure and we set the most stringent limit on this decay.  Thanks to the simultaneous readout of the heat and light signals we reach the lowest background level ever achieved with bolometric experiments: (3.6$^{+1.9}_{-1.4}$)$\times$10$^{-3}$\ckky.  

Although a detailed discussion of the background components is the subject of a dedicated future paper, it is worth stressing that we expect contributions at the level of $\sim$ 10$^{-3}$ \ckky\ both from cosmic muon induced events and  from contaminations of the cryogenic setup which hosted CUORE-0 \cite{Alduino:2016vtd}.
The successful operation of  CUPID-0 and the capability to reject the \al\  induced background is a key milestone for the next-generation tonne-scale project CUPID. 

This work was partially supported by the European Research Council (FP7/2007-2013) under Low-background Underground Cryogenic Installation For Elusive Rates Contract No. 247115. We are particularly grateful to M. Iannone for the help in all the stages of the detector construction, A. Pelosi for the construction of the assembly line, M. Guetti for the assistance in the cryogenic operations, R. Gaigher for the calibration system mechanics, M. Lindozzi for the development of cryostat monitoring system, M. Perego for his invaluable help,  the mechanical workshop of LNGS (E. Tatananni, A. Rotilio, A. Corsi, and B. Romualdi) for the continuous help in the overall setup design. A.~S.~Z. is supported by the Initiative Doctorale Interdisciplinaire 2015 project funded by the Initiatives d?excellence Paris-Saclay, ANR-11-IDEX-0003-0. We acknowledge the Dark Side Collaboration for the use of the low-radon clean room.
This work makes use of the DIANA data analysis and APOLLO data acquisition software which has been developed by the CUORICINO, CUORE, LUCIFER, and CUPID-0 Collaborations.
\bibliography{main}

\end{document}